\documentclass[aps,prx,reprint,superscriptaddress,showpacs]{revtex4-1}
\usepackage{ifpdf}
\usepackage{color}
\usepackage{graphicx}
\usepackage[american]{babel}
\usepackage[utf8]{inputenc}
\usepackage[T1]{fontenc}
\usepackage{bbm,ae,aecompl}
\usepackage{amsmath}
\usepackage{amssymb}
\usepackage{amstext}
\usepackage{amsthm}
\usepackage{latexsym}
\usepackage{verbatim}
\usepackage{natbib}
\usepackage{array}

\usepackage{ulem}   

\begin{document}

\title{ARPES view of orbitally resolved quasiparticle lifetimes in iron pnictides}
\date{\today}

\author{Véronique Brouet}
\author{David LeBoeuf}

\author{Ping-Hui Lin}

\author{Joseph Mansart}
\affiliation{Laboratoire de Physique des Solides, Universit\'{e} Paris-Sud, UMR 8502, B\^at. 510, 91405 Orsay, France}

\author{Amina Taleb-Ibrahimi}

\author{Patrick Le F\`evre}

\author{François Bertran}
\affiliation{Synchrotron SOLEIL, L'Orme des Merisiers, Saint-Aubin-BP 48, 91192 Gif sur Yvette, France}



\author{Anne Forget}

\author{Dorothée Colson}
\affiliation{Service de Physique de l'Etat Condens\'{e}, Orme des Merisiers, CEA Saclay, CNRS-URA 2464, 91191 Gif sur Yvette Cedex, France}

\begin{abstract}
We study with ARPES the renormalization and quasiparticle lifetimes of the $d_{xy}$ and $d_{xz}$/$d_{yz}$ orbitals in two iron pnictides, LiFeAs and Ba(Fe$_{0.92}$Co$_{0.08}$)$_2$As$_2$ [e.g. Co8]. We find that both quantities depend on orbital character rather than on the position on the Fermi Surface (for example hole or electron pocket). In LiFeAs, the renormalizations are larger for $d_{xy}$, while they are similar on both types of orbitals in Co8. The most salient feature, which proved robust against all the ARPES caveats we could think of, is that the lifetimes for $d_{xy}$ exhibit a markedly different behavior than those for $d_{xz}$/$d_{yz}$. They have smaller values near $E_F$ and exhibit larger $\omega$ and temperature dependences. While the behavior of $d_{xy}$ is compatible with a Fermi liquid description, it is not the case for $d_{xz}$/$d_{yz}$. This situation should have important consequences for the physics of iron pnictides, which have not been considered up to now. More generally, it raises interesting questions on how a Fermi liquid regime can be established in a multiband system with small effective bandwidths. 
\end{abstract}

\maketitle

\begin{figure*}[htb]
\centering
\includegraphics[width=1\textwidth]{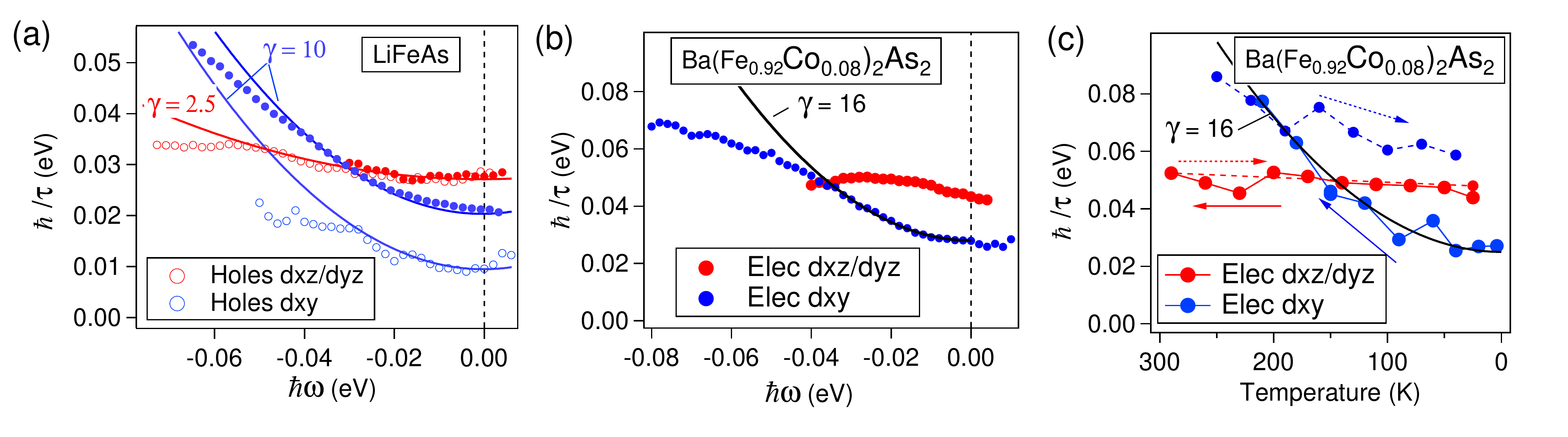}
\caption{Lifetime $\hbar$/$\tau$ (eV) extracted from ARPES linewidths (see section V) in (a) LiFeAs for hole and electron pockets at 25K as a function of binding energy $\hbar\omega$ (b) Co8 for electron pockets at 25K as a function of binding energy $\hbar\omega$ (c) Co8 for electron pockets at $E_F$ as a function of temperature. As we could never obtain a behavior reversible in temperature for $d_{xy}$, we indicate width measured when warming up (large symbols) and cooling down (small symbols). The lines are fits to simple Fermi liquid behaviors, as discussed in section V-B. They are calculated from Eq. 1, with the values of $\gamma$ indicated on the figure and adequate offsets.}
\label{Lifetimes}
\end{figure*}

\section{Introduction}
The discovery of iron-based superconductors in 2008 has brought to physicists both a new family of high temperature superconductors and a new type of correlated systems \cite{ChubukovPhysToday15}. They have a multiband electronic structure, where the three Fe $3d$ $t_{2g}$ orbitals cross the Fermi level and form small hole and electron pockets. This is unusual among correlated systems and it has been argued that the value of Hund\rq{}s couplings between electrons in different orbitals, rather than simply the Coulomb repulsion, controls the strength of electronic correlations in these systems \cite{HauleNJP09}. These correlations may change significantly for different families or different orbitals, $d_{xy}$ being more correlated in many cases \cite{YinNatureMat11,deMediciPRL14}. The role of the different orbitals in the electronic properties is a major issue. Angle resolved photoemission (ARPES) should be an ideal tool to discuss these questions, as it has a unique ability to probe the properties of each band and/or orbital individually. There has been a huge effort to map the electronic structure in different families of iron pnictides with ARPES and compare the results with band structure calculations \cite{RichardReports11,KordyukReview13,YeFengPRX14}. However, very little is known about the quasiparticles (QP) lifetimes $\tau$ of the different orbitals, which could, in principle, be extracted from the ARPES linewidth $\delta\nu\propto1/\tau$ \cite{DamascelliRMP2003,VallaPRL99}. Despite a few studies \cite{KordyukPRB11,YeFengPRX14,Fink15}, there are no systematic investigation of ARPES linewidths for the different bands of one system. It is often assumed that the lines are too broad - especially in the case of Co doped BaFe$_2$As$_2$ \cite{VanHeumenPRL11,WadatiPRL10} - to perform such analysis. In this paper, we argue that meaningful information can be extracted from the ARPES linewidths and that they reveal an intriguing and universal difference in behavior between $d_{xy}$ and $d_{xz}$/$d_{yz}$ orbitals.

Whether the metallic behavior in iron pnictides can be described within a Fermi liquid (FL) framework is an essential question. The QP lifetimes would then obey the following formula.
\begin{equation}
\hbar/\tau = \gamma\left((\hbar\omega)^2+(\pi k_BT)^2\right)
\label{Eq1}
\end{equation}
However, it is noteworthy that the resistivity of  BaFe$_2$As$_2$ becomes linear near 8\% Co doping \cite{FangMazinPRB09,RullierAlbenquePRL09,ChuFisherPRB09}, deviating from the canonic T$^2$ dependence of a FL. This has often been taken as a sign of non FL behavior, bearing analogies with cuprates \cite{CooperScience09} or organic superconductors \cite{DoironPRB09}. However, the interpretation of this linear resistivity is not straightforward, due to the contribution of holes and electrons with different orbital characters that could have different characteristics. Indeed, Rullier-Albenque \textit{et al.} argued that the scattering rates remain proportional to T$^2$ in the entire Co doping range of BaFe$_2$As$_2$ \cite{RullierAlbenquePRL09}. This is precisely where ARPES could help clarifying the situation by resolving directly the contribution of each type of carrier. 

\vspace{3mm }

To perform our study, we take two examples : LiFeAs and BaFe$_2$As$_2$ doped with 8\% Co (called Co8 in the following). We choose them because they are both metals down to relatively low temperatures and both become superconducting below 18K for LiFeAs \cite{TappPRB08} and 22K for Co8 \cite{NiPRB08}. On the other hand, LiFeAs appears slightly more correlated in DMFT calculations \cite{YinNatureMat11,FerberPRB12}. This is mainly due to the elongation of the Fe tetrahedra along z, which tends to reduce the hopping between neighboring Fe and enhance correlations, especially for the in-plane dxy orbital \cite{YinNatureMat11}. Moreover, the two compounds have slightly different structures, with different stackings of the FeAs slabs \cite{ChubukovPhysToday15}, which introduces differences in their electronic structure that are useful to discuss the intrinsic origin of the behaviors we observe. In particular, there is no disorder in LiFeAs introduced by Co doping and the surface is non-polar, suppressing some of the problems discussed in ref. \cite{VanHeumenPRL11,WadatiPRL10}. The general electronic structure has already been studied in some details by ARPES, both for Co8 \cite{ZhangPRB11,BrouetPRL13,RichardReports11,MalaebJPhysSocJapan09} and LiFeAs \cite{BorisenkoLiFeAs,KordyukPRB11,LeePRL12}. We add here more focused information on the separation of hole and electron bands of different orbital character and their linewidths. 

In Fig. \ref{Lifetimes}, we summarize the main results of this paper. In panel (a), the evolution of the lifetime in LiFeAs is shown as a function of binding energy $\hbar\omega$ for the $d_{xz}$/$d_{yz}$ (red) and $d_{xy}$ (blue) orbitals on hole (open symbols) and electron (closed symbols) pockets. In panel (b), the same information is given for electron pockets in Co8. We do not give results for the hole pockets in Co8, because they significantly overlap and are more difficult to resolve from each other \cite{BrouetPRL13,ZhangPRB11}. In panel (c), we further show the temperature dependence of the lifetime at the Fermi level in Co8. There are two striking effects. First, the $\hbar\omega$ dependence of the linewidths in LiFeAs are very similar for the hole and electron bands of the same orbital character. This evidences that the orbital character rather than the hole/electron character or the location on Fermi Surface (FS) determines the lifetime behavior. Second, there is a contrast in the behavior of $d_{xz}$/$d_{yz}$ and  $d_{xy}$. There is a clear dependence as a function of $\hbar\omega$ for $d_{xy}$ below 50meV that is missing or very reduced for $d_{xz}$/$d_{yz}$. The fact that the temperature dependence is similarly larger on $d_{xy}$ compared to $d_{xz}$/$d_{yz}$ in Co8 [Fig. \ref{Lifetimes}(c)], following expectations from Eq. 1, reinforces the idea that the difference is intrinsic. Moreover, the absolute values at the Fermi level $E_F$ are generally larger for $d_{xz}$/$d_{yz}$ than $d_{xy}$.  As the same tendencies are detected for LiFeAs and Co8, this appears as a rather universal feature, which likely has important consequences for the physics of iron pnictides. 

In this paper, we detail the different steps necessary to obtain the data in Fig. 1. We first locate each band on the FS (section III). This shows that their properties (dispersion and width) can indeed be very clearly resolved from each other. We extract effective masses by comparison to band calculations. They are enhanced by a factor $\sim$ 2 in Co8 for all bands and of $\sim$ 2 for $d_{xz}$/$d_{yz}$ and 3.5 for $d_{xy}$ in LiFeAs (see Table I). This confirms predictions from DMFT \cite{YinNatureMat11,FerberPRB12} that the orbitals start to differentiate in LiFeAs, $d_{xy}$ being more renormalized. One potential problem in determining lifetimes from ARPES is the three dimensionality (3D) of the electronic structure, which can induce extrinsic broadening \cite{StrocovJEleSpec03,BansilPRB05}. In section IV, we show that 3D effects are present in both compounds, but they are quite different, due to the different stacking of the FeAs slabs, so that we can rule out such an effect as the origin of the difference in $d_{xy}$ and $d_{xz}$/$d_{yz}$ linewidths. Finally, we return to the discussion of the lifetimes in section V. We first discuss the absolute values of the linewidths, including the role of finite resolution and impurities. We then detail how the linewidths, determined by fitting Momentum Distribution Curves (MDC) in \AA$^{-1}$, are converted to lifetimes in eV by using the slope of the band dispersion. We conclude that the difference in behavior of $d_{xy}$ and $d_{xz}$/$d_{yz}$ is robust against all these problems.

\vspace{3mm}
\begin{table*}[tb]
\caption{\label{''Table 1''} First and second columns : renormalization and shift needed to fit the experimental dispersion with the theoretical ones (see Fig. 2 and 3). Third column : energy position with respect to E$_F$ of the top or bottom of the band (for hole bands, we use the position of the fitted calculation). Fourth and fifth columns : number of carriers $n$ per Fe in each band, after integration over $k_z$, from theory and experiment.}

\begin{tabular}{|m{3.5cm}|>{\centering\arraybackslash}m{2.3cm}|>{\centering\arraybackslash}m{2.3cm}|>{\centering\arraybackslash}m{2.3cm}|>{\centering\arraybackslash}m{2.3cm}|>{\centering\arraybackslash}m{2.3cm}|}

\hline
                                          &    Renormalization       &      Shift (meV)    &     Band extremum (meV)   &   $n/Fe$ (th.)   & $n/Fe$ (exp.) \\
\hline
Co8 &    &    &    &    & \\    
\hline
   electron $d_{xz}$/$d_{yz}$ &           2.3          &     100   &   40                     &       0.1                         &          0.04  \\
   electron $d_{xy}$               &          2.1         &   100  &   150           &       0.1   & 0.06                                    \\

  \hline
  LiFeAs &    &    &    &     &   \\
\hline
     electron $d_{xz}$/$d_{yz}$ &     2.2                 &   0 &  55                         &    0.18                              &         0.1 \\
 hole $d_{xz}$/$d_{yz}$ &           1.8          &  -70   &    20 & 0.09 &0.02               \\  
   electron $d_{xy}$           &           4         &   -20  &   150                    &      0.08                           &       0.1  \\
 hole $d_{xy}$               &            3.3   & 35     &    80                      &          0.17                                    &                    0.21  \\

   \hline
\end{tabular}
\end{table*}

\section{Experimental details}
Single crystals were grown using a FeAs self-flux method and were investigated in details by transport measurements \cite{RullierAlbenquePRL09,RullierAlbenquePRL12}. ARPES experiments were carried out at the CASSIOPEE beamline at the SOLEIL synchrotron, with a Scienta R4000 analyser, an angular resolution of $0.3^{\circ}$ (0.015 \AA$^{-1}$ at 34eV) and an energy resolution better than $15$ meV. For these measurements, the linearly polarized light was set in the sample plane along a Fe-Fe bond [$\Gamma$X direction of the Brillouin Zone (BZ)]. In this configuration, polarization selection rules \cite{DamascelliRMP2003,BrouetPRB12} select orbitals \textit{even} with respect to the plane containing this axis and the normal to the sample surface and \textit{odd} with respect to the plane perpendicular to the previous one. Note that the band parity may be different from its main orbital character, because there are 2 inequivalent Fe in the unit cell \cite{BrouetPRB12,MoreschiniPRL14}. The photon energy selects a particular $k_z$ value, which we estimate through the free electron final state approximation, already heavily used in iron pnictides \cite{VilmercatiPRB09,BrouetPRB09,MalaebJPhysSocJapan09}. 
\begin{equation}
k_z=\sqrt{2m/\hbar^2*(h\nu-W+V_0)-k_{//}^2}
\label{Eq2}
\end{equation} 
For the inner potential V$_0$, we use 11eV for LiFeAs and 14eV for Co8 (see section IV). Note that because Co8 has a body centered structure \cite{ChubukovPhysToday15}, two adjacent BZ are shifted in $k_z$, which is not the case in LiFeAs. 

 Band structure calculations were perfomed within the local density approximation (LDA), using the Wien2K package \cite{Wien2k}, with the experimental structures. The Co doping of 8\% was treated in the virtual crystal approximation \cite{BrouetPRL13}.

\section{Band dispersion}

\begin{figure}[b]
\centering
\includegraphics[width=0.5\textwidth]{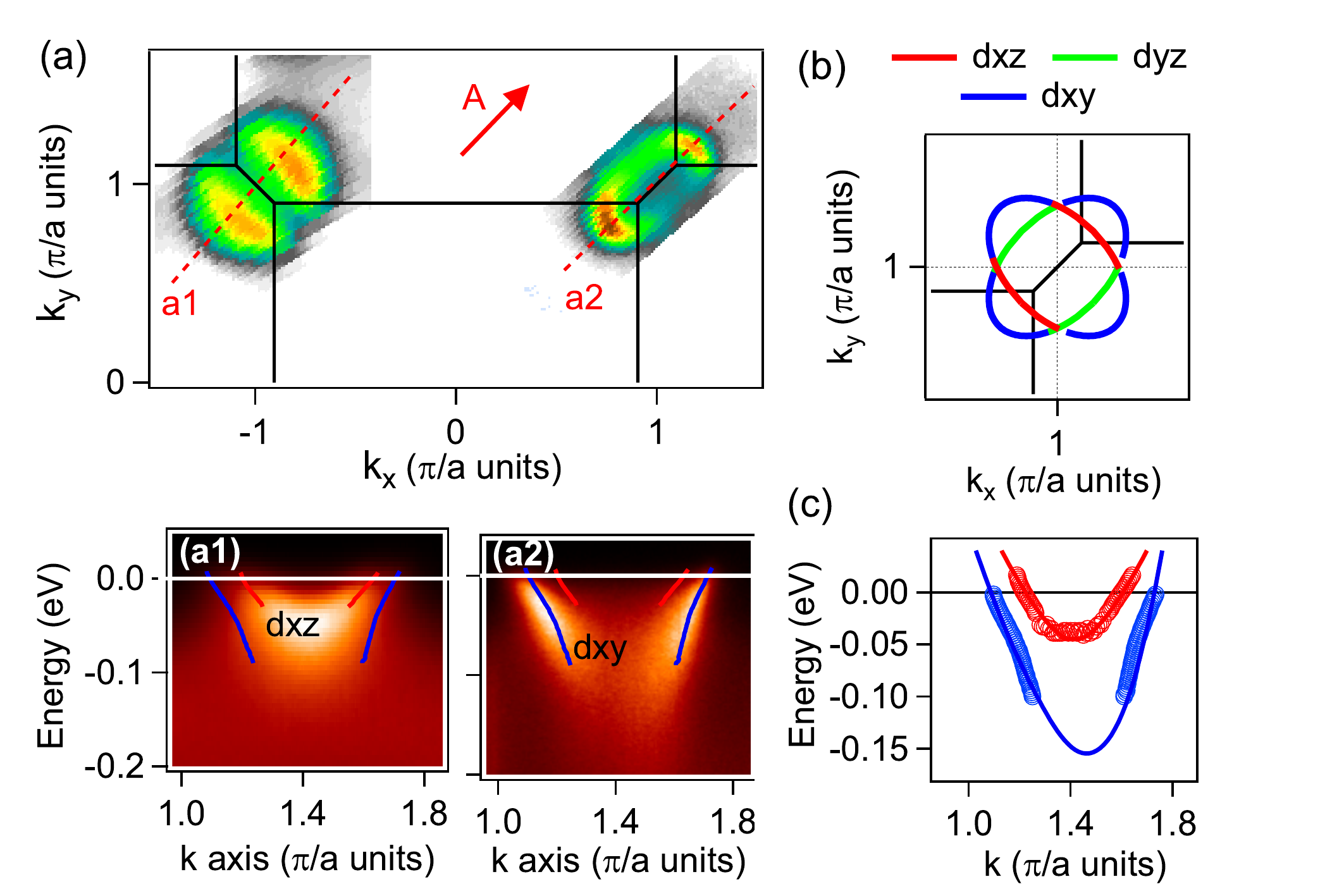}
\caption{(a) Fermi Surface measured in Co8 at T=25 K, with polarization $A$ along the red arrow and photon energy 34 eV. (a1)-(a2) Energy-momentum images of the two different cuts indicated in (a). (b) Sketch of the orbital characters on the two electron pockets expected by calculation. (c) Experimental band dispersion (symbols) compared with calculated bands shifted and renormalized as indicated in Table 1.}
\label{Co8_FS_Disp}
\end{figure}

In Fig. \ref{Co8_FS_Disp}, we present the ARPES measurements of the $d_{xz}$/$d_{yz}$ and $d_{xy}$ parts of the electron pockets in Co8. A sketch of the distribution of orbital character on the electron pockets expected in LDA calculations is recalled in Fig. \ref{Co8_FS_Disp}(b). We have explained in \cite{BrouetPRB12}, which bands can be observed in this experimental configuration and why. Along (a1), we observe the shallow $d_{xz}$/$d_{yz}$ band, while along (a2), only the deeper $d_{xy}$ band is detected. Their very different dispersion makes it easy to ensure that only one band is observed in each case. This difference in dispersion is expected from band structure calculations. In Fig. \ref{Co8_FS_Disp}(c), we show the experimental dispersions extracted by MDC analysis (symbols), together with the calculated band dispersions renormalized by $\sim$ 2 and shifted up (see exact values in Table I). The colors correspond to the main orbital character. The dispersion is calculated exactly at the experimental $k_z$, obtained using Eq. 2. The asymmetry of the $d_{xy}$ dispersion is due to the change of $k_z$ along the cut (through $k_{//}$) and is well reproduced in the calculation. To match the data, it is however further necessary to shift up the bands by 0.1eV before renormalizing them. This corresponds to a \lq\lq{}shrinking\rq\rq{} of the electron pockets that we described in details before \cite{BrouetPRL13}. A concomitant downward shift of the hole bands allows charge conservation.

\begin{figure*}[tb]
\centering
\includegraphics[width=1\textwidth]{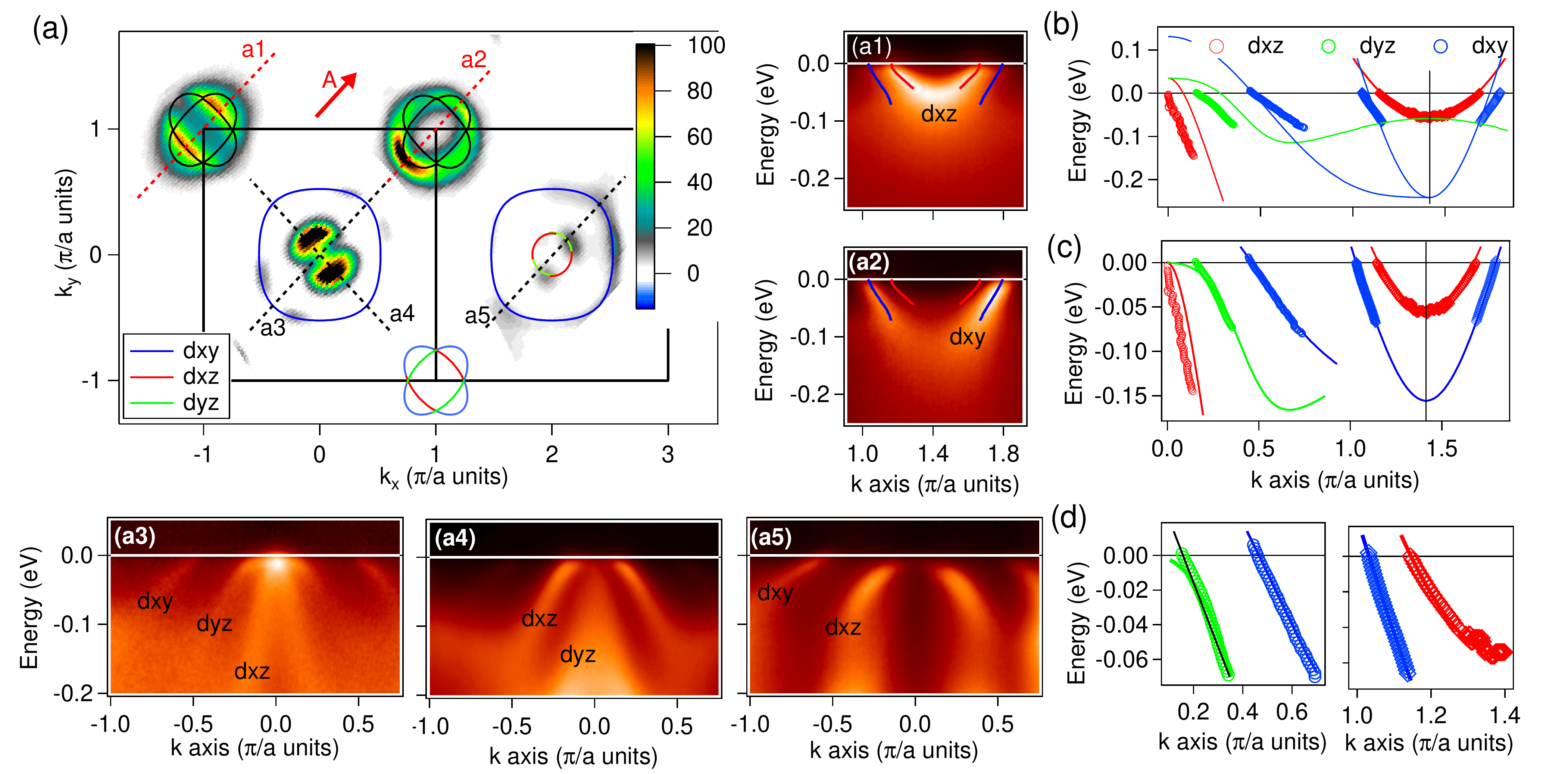}
\caption{(a) Fermi Surface measured in LiFeAs at T=25 K, with polarization $A$ along the red arrow and photon energy 34eV. Sketches of the different pockets with orbital characters are indicated as guides to the eye. (a1)-(a5) : energy-momentum images of the different cuts indicated on the FS map.  (b) Experimental dispersions of all bands at $k_z$=0 (points). Calculated dispersions renormalized by a factor 2 (lines). (c) Same experimental dispersions with calculation shifted and renormalized as indicated in Table 1. (d) Zoom on the dispersion of the different bands near $E_F$. Color lines are the LDA dispersion of (c) and black line on $d_{yz}$ hole a linear fit.}
\label{LiFeAs_FS_Disp}
\end{figure*}

In Fig. \ref{LiFeAs_FS_Disp}, we present the electronic structure of LiFeAs, measured in similar experimental conditions. Again, we can distinguish very well the $d_{xz}$/$d_{yz}$ and $d_{xy}$ parts of the electron pockets in panels (a1) and (a2). Although the two bands are very clear in this geometry, it was not the case in previous studies, where the $d_{xy}$ band was either completely missing \cite{LeePRL12} or observed simultaneously with $d_{xz}$/$d_{yz}$ \cite{BorisenkoLiFeAs}. Three different bands form the hole pockets at the zone center, as detailed in panels (a3) to (a5). The most outer hole band has $d_{xy}$ character and forms a large squarish pocket [blue contour in Fig. \ref{LiFeAs_FS_Disp}(a)]. The  $d_{xz}$ and  $d_{yz}$ bands form smaller hole pockets around the zone center, one having roughly a circular contour and one barely crossing the Fermi level. This is in good agreement with previous ARPES measurements \cite{BorisenkoLiFeAs,LeePRL12}. A big advantage of LiFeAs is that all these hole bands are well separated and each of them can be unambiguously attributed to one orbital. Therefore, we have a unique opportunity here to study in details the contribution of  $d_{xz}$/$d_{yz}$ and $d_{xy}$, both for hole and electron pockets.

In Fig. \ref{LiFeAs_FS_Disp}(b), we extract the dispersions of these different bands at $k_z$=0 (we choose the appropriate photon energy in accordance to section IV). We compare them with the calculated bands divided by a factor 2, this value being chosen to get a global view of the situation. Obviously, the hole $d_{xz}$/$d_{yz}$ bands are shifted down compared to this calculation. On the contrary, the electron bands and the hole $d_{xy}$ band are not significantly shifted from calculation. This contrasts with the global shrinking discussed before in Co8. Here, holes are mainly transferred from $d_{xz}$/$d_{yz}$ to $d_{xy}$ (see table I and section IV for more details). This transfer is induced by correlations and was correctly predicted by DMFT \cite{YinNatureMat11} and already observed by ARPES \cite{BorisenkoLiFeAs,LeePRL12}. This puts $d_{xy}$ closer from half-filling, which is one of the reasons why it is predicted to get more correlated \cite{deMediciPRL14}.

In Fig. \ref{LiFeAs_FS_Disp}(c), we fit each band individually by adjusting the shift and renormalization with parameters reported in Table 1. As was already clear in Fig. \ref{LiFeAs_FS_Disp}(b), $d_{xy}$ disperses more slowly than the calculation divided by 2, both for hole and electron pockets, meaning some additional renormalization is needed. We find values of 3.3 and 4 for $d_{xy}$, compared to $\sim$2 for $d_{xz}$/$d_{yz}$. This shows that the renormalization value is fixed by orbital character rather than hole or electron character. In LiFeAs, $d_{xy}$ starts to differentiate from other orbitals by exhibiting larger correlation effects. In Fig. \ref{LiFeAs_FS_Disp}(d), we zoom on the dispersion of each band to show in more details the quality of the fit at low energies.

\section{Three dimensional dispersion}

In a 3D system, the ARPES spectra are often broadened by the photoemission process \cite{StrocovJEleSpec03,BansilPRB05} and their linewidths may not reflect the QP lifetimes anymore. The short electron escape depth $\lambda$ in the photon energy range we use (20-100eV) leads to an integration over some finite $k_z$ range ($\delta$$k_z$$\propto$1/$\lambda$). This can induce severe broadening of the ARPES spectra if the $k_z$ dispersion is large. In this section, we pay particular attention to this point to understand how it could affect our linewidth analysis.

\begin{figure}[tb]
\centering
\includegraphics[width=0.5\textwidth]{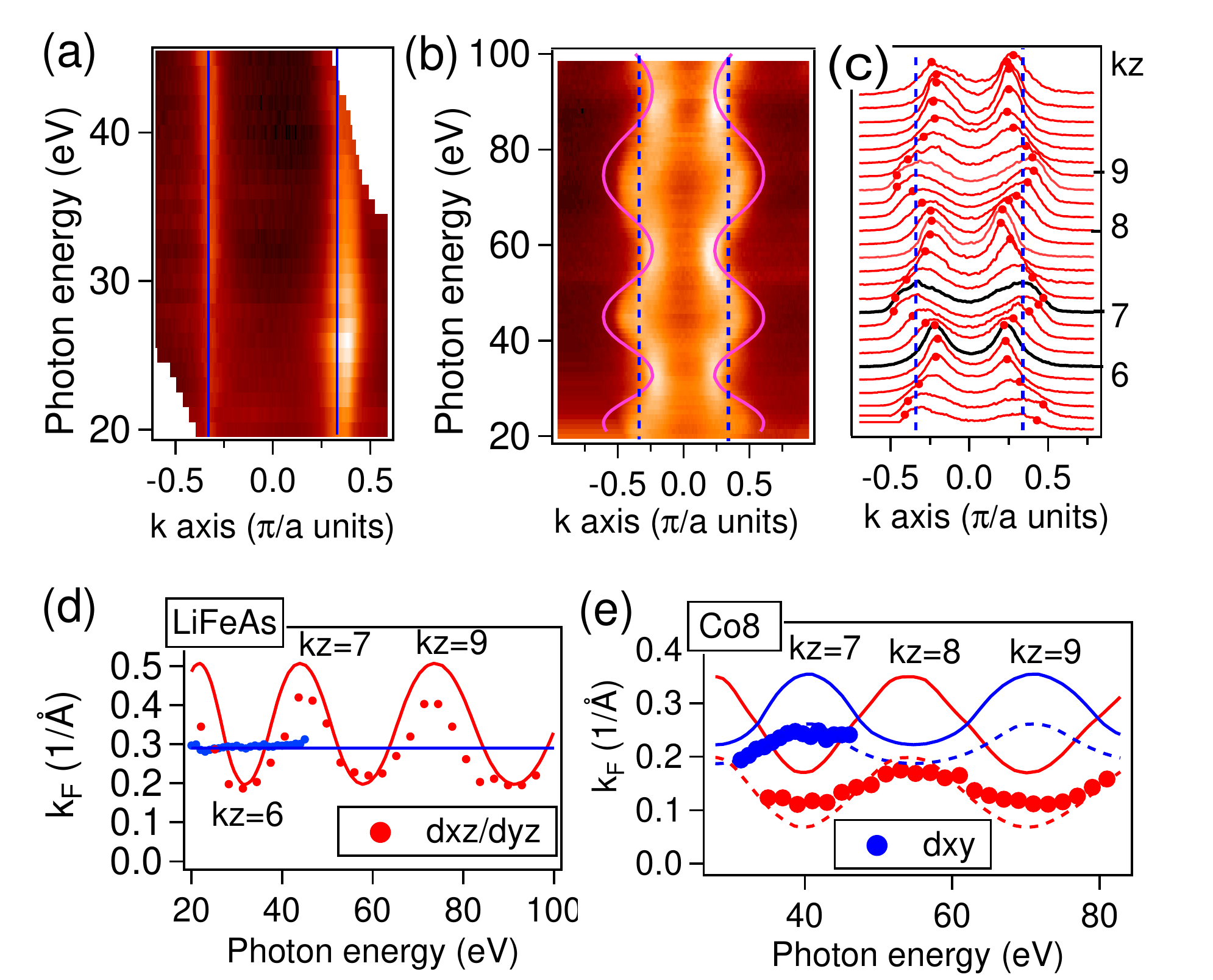}
\caption{(a) Spectral weight in LiFeAs integrated in a 5meV window around E$_F$ for the $d_{xy}$ electron band as a function of photon energy (cut (a2) in Fig. 3). Lines are theoretical variation for k$_F$. (b) Same for the $d_{xz}$/$d_{yz}$ electron band (cut (a1) in Fig. 3). Dotted line is the position of $d_{xy}$. (c) Spectra corresponding to (b). (d) Experimental (points) and calculated (lines) variations of k$_F$ with photon energy in LiFeAs. (e) Same as (d) for Co8. The dotted lines indicate $k_F$ calculated after a shift of 0.1eV. }
\label{3Deffects}
\end{figure}

Fig. \ref{3Deffects} shows maps of the spectral weight in LiFeAs for $d_{xy}$ (a) and $d_{xz}$/$d_{yz}$ (b) electron bands at $E_F$, as a function of photon energy or equivalently of $k_z$ using Eq. 2. No variation of $k_F$ is detected for $d_{xy}$, while it is clear and periodic for $d_{xz}$/$d_{yz}$ (note the different photon energy ranges). This is in good agreement with calculations (solid lines) for which $d_{xy}$ is strictly 2D, while $d_{xz}$/$d_{yz}$ exhibits clear 3D dispersion. The experimental 3D dispersion is not as strong for $d_{xz}$/$d_{yz}$ as predicted in the calculation. In fact, the corresponding spectra [Fig. \ref{3Deffects} (c)] are good lorentzians only near $k_z$=0 (where we performed most of our analysis) and broaden asymmetrically towards $k_z$=1 (see black spectra for $k_z$=6 and 7). Assuming this distortion of lineshape is due to some averaging over $k_z$, we extracted the variation of $k_F$ with $k_z$ using the outer edge of the spectra, as shown by the red points. The two dispersions of the electron bands are then summarized in Fig. \ref{3Deffects}(d). The dispersion extracted in this way is 30\% smaller than the theoretical one. From these dependences, we can extract the number of electrons in each band. Assuming the electron pockets can be divided into a squarish inner sheet of $d_{xz}$/$d_{yz}$ symmetry and an outer sheet of $d_{xy}$ symmetry, we obtain 0.1 electron in each band after integrating over $k_z$. These numbers are compared in Table I with the theoretical ones. The number of holes, for which we did not measure significant 3D dispersion at $E_F$, are also added to this table. We find number of carriers reduced by $\sim$10-20\% compared to calculations and compatible with charge neutrality.

We reported the 3D electronic structure of Co8 in \cite{BrouetPRB12} and just recall the results in Fig. \ref{3Deffects}(e). The different stacking of the FeAs slabs in Co8 yields quite different 3D effects, with a sizable energy dependence of $d_{xy}$, which is nearly as large as that of $d_{xz}$/$d_{yz}$ in theory (solid lines, the maxima are shifted in $k_z$, because the two bands are measured in two adjacent BZ, where $k_z$ is shifted by 1). Our results are quite well reproduced by calculations, both for $d_{xy}$ and $d_{xz}$/$d_{yz}$, if we compute $k_F$ at -0.1eV (dotted line) rather than $E_F$ (solid line), to take into account the FS shrinking. The variation for $d_{xz}$/$d_{yz}$ is again slightly reduced in experiment. The number of electrons is divided by 2 compared to the theoretical expectation. This very large shrinking in Co8, almost absent in LiFeAs, is an intriguing feature that should be further explored. It was assigned by Ortenzi \textit{et al.} to interband interactions mediated by spin fluctuations \cite{OrtenziPRL09} and could then suggest a different role of such fluctuations in the two compounds. 

These differences in $k_z$ variations turn out to be useful to estimate the importance of 3D effects in the measured linewidth. In LiFeAs, one could suppose at first that $d_{xy}$ is narrower because it is more 2D. However, $d_{xy}$ is nearly as 3D as $d_{xz}$/$d_{yz}$ in Co8, and yet, it is also narrower than $d_{xz}$/$d_{yz}$. Moreover, the $k_z$ dispersion is nearly twice smaller for $d_{xz}$/$d_{yz}$ in Co8 compared to LiFeAs, but the linewidths are broader (see Fig. 1). This suggests that the contribution of 3D effects in linewidths does not dominate. Similarly, the $d_{xz}$/$d_{yz}$ hole band in LiFeAs has a much smaller $k_z$ dispersion than the electron band (a factor 7 smaller in theory), but a similar linewidth. We conclude that the difference in linewidth behavior between $d_{xz}$/$d_{yz}$ and $d_{xy}$ cannot be attributed to $k_z$ dispersion.

\section{Linewidth analysis}

\subsection{Linewidth at the Fermi level}
\begin{figure}[tbp]
\centering
\includegraphics[width=0.45\textwidth]{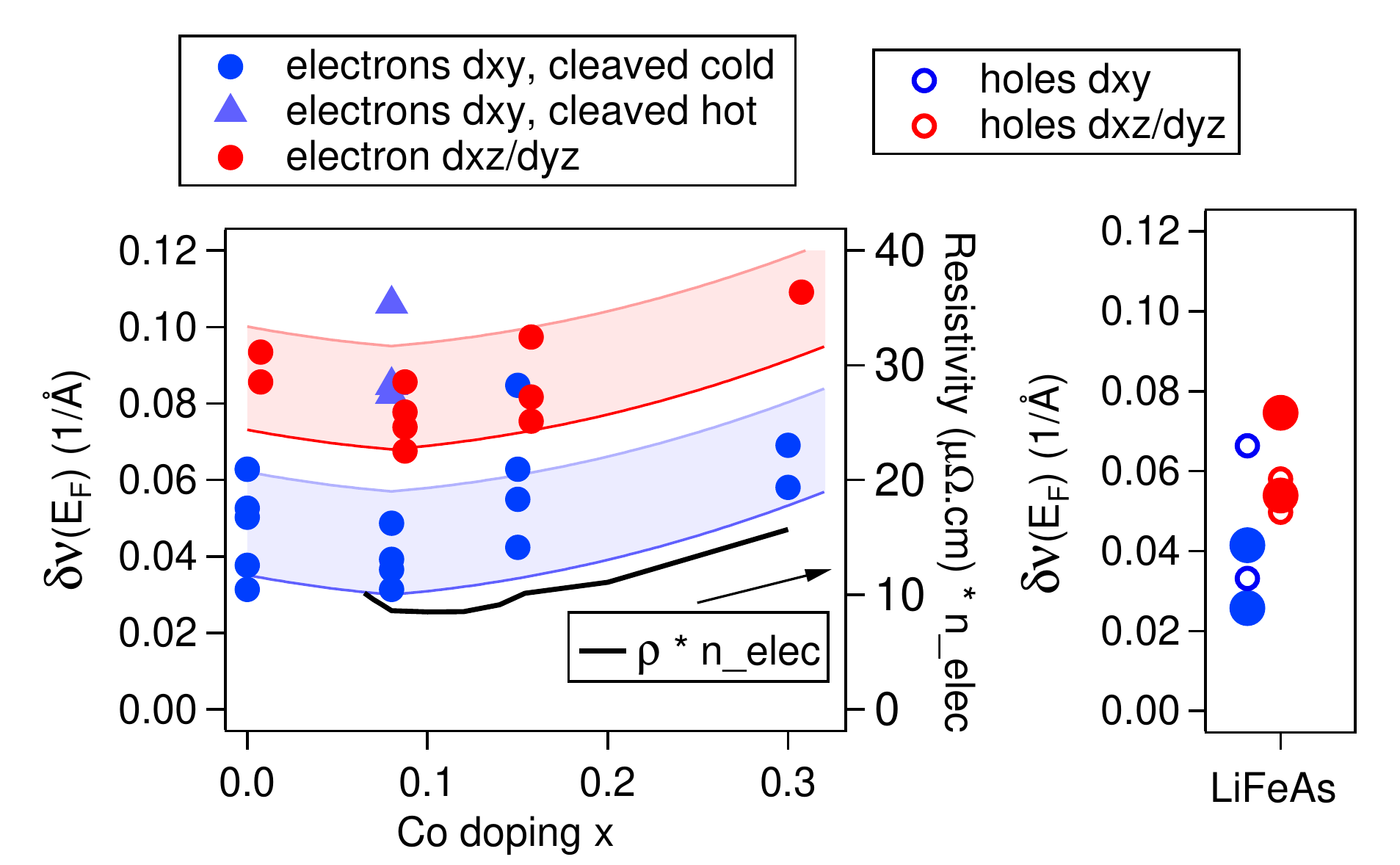}
\caption{ MDC half width at half maximum $\delta\nu$ observed at $E_F$ and low temperatures (25-40K, except BaFe$_2$As$_2$, 150K) in different samples of Ba(Fe$_{1-x}$Co$_x$)$_2$As$_2$ (left) and LiFeAs (right). Black line (right axis of Co panel) : residual resistivity in the normal state taken from ref. \cite{RullierAlbenquePRL09} and multiplied by the number of electrons from ref. \cite{BrouetPRL13}.}
\label{MeanFreePath}
\end{figure}

We now return to the discussion of Fig. \ref{Lifetimes}. We consider first the different values at the Fermi level for $d_{xz}$/$d_{yz}$ and $d_{xy}$. The meaning of these differences may better appear in Fig. \ref{MeanFreePath}, where we have gathered values of MDC linewidths measured in different samples, different cleaves and/or different Co contents. This reveals quite a large scattering in absolute values, usually by a factor 2 and even much larger for samples cleaved at high temperatures (blue triangles). Some trend remain however clear, such as the smaller values for $d_{xy}$ compared to $d_{xz}$/$d_{yz}$ or the increase of both linewidths as a function of Co doping. We do not observe a larger increase of $d_{xy}$ linewidth with Co content in BaFe$_2$As$_2$, contrary to what was recently reported in Co-doped LiFeAs \cite{YeFengPRX14}. On the other hand, $d_{xy}$ was found to be more sensitive to temperature cycles (see Fig. 1), which we attribute to the effect of impurities formed during such cycles. This is consistent with the idea of a larger or different coupling of $d_{xy}$ to impurities. Interestingly, the trend of the variations with Co content scales very well with expectations deduced from residual resistivity \cite{RullierAlbenquePRL09}. Expecting $\rho\propto1/n\tau$, we plot on the right axis the residual resistivity multiplied by the number of electrons estimated in  \cite{BrouetPRL13} (black line, we neglected the hole contribution, as suggested in ref. \cite{RullierAlbenquePRL09}). The similar tendency is a very compelling indication that the linewidths indeed reflect intrinsic physics of the compounds, even if there may be an additional constant term of extrinsic origin. 
 
The linewidths in LiFeAs are somewhat smaller than the Co ones, but not dramatically. For the hole $d_{xy}$ band, our value is similar to that of the previous report by Kordyuk \textit{et al.} \cite{KordyukPRB11}. In fact, our data do not support the common idea that ARPES spectra are much broader in Co-doped BaFe$_2$As$_2$ than LiFeAs. To explain the large scattering in the absolute values of the linewidths, we infer that they contain a rather large contribution from impurities mainly created at the surface during the cleave. This also explains the much smaller values of 1/$\tau$ near $E_F$ measured by optics, namely 7~meV for Co8 \cite{TytarenkoCondMat15} and 3~meV for LiFeAs \cite{DaiPRX15}. However, this impurity contribution should be independent of binding energy and temperature \cite{VallaPRL99}, so that it can be considered as a mere offset. Indeed, we usually observe similar energy dependences of linewidths for samples exhibiting different absolute values. There is also a constant contribution from experimental resolution that we have not attempted to subtract. 

Transport \cite{RullierAlbenquePRL09} and quantum oscillation experiments \cite{ColdeaPRL08,PutzkePRL12} have often suggested longer mean free paths for electrons than holes. This is not corroborated by these measurements, as we observe similar linewidths for hole and electron pockets for a given orbital. Nevertheless, as the orbital content is different for hole and electron pockets, taking into account different behaviors for $d_{xz}$/$d_{yz}$ and $d_{xy}$ may reconcile these observations. 

\subsection{Dependence with binding energy and temperature}
\begin{figure}[t]
\centering
\includegraphics[width=0.45\textwidth]{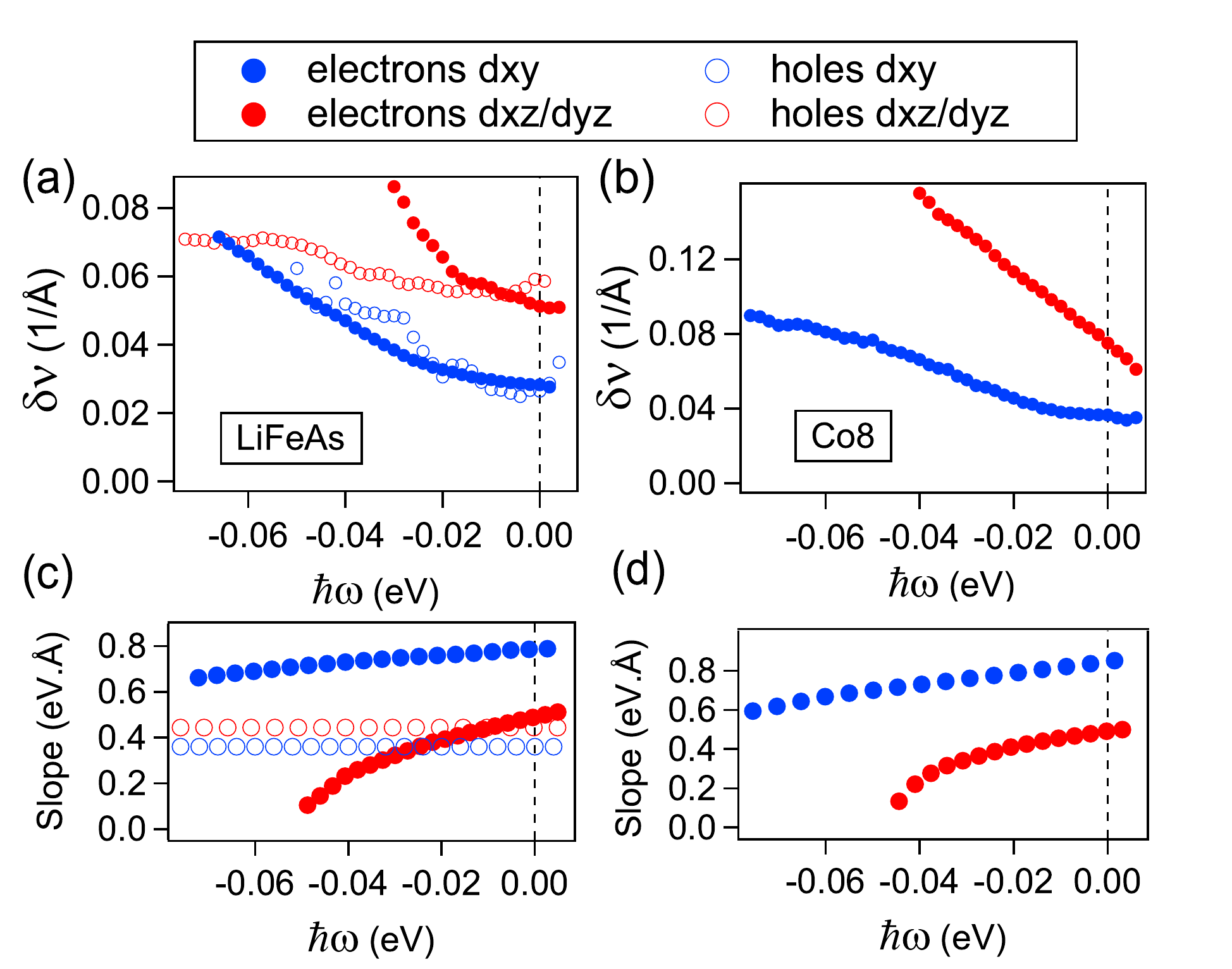}
\caption{ (a-b) MDC half width at half maximum $\delta\nu$ as a function of binding energy in LiFeAs (a) and Co8 (b). (c-d) Slopes of the dispersion used to compute 1/$\tau$ (Eq. 3) in LiFeAs (c) and Co8 (d). For electron bands, it is the slope of the renormalized LDA dispersion. For hole bands, it is straight lines, as they give a better or equivalent fit than the LDA (see Fig. 3d). }
\label{Largeur_MDC}
\end{figure}

The linewidths are extracted experimentally by fitting the MDCs to lorentzians. This is much easier than fitting directly the Energy Distribution Curves (EDC), whose background is more difficult to model \cite{DamascelliRMP2003,VallaScience99}. These MDC linewidths, in units of  \AA$^{-1}$, depend both on band properties (such as v$_F$) and many body effects. To isolate the contribution of many-body effects, we now compute 1/$\tau$ by multiplying the MDC linewidth by the slope of the dispersion. As long as the experimental dispersion is proportional to the theoretical one ($v_{bare}$), one can define a constant renormalization value $Z=v_F/v_{bare}$ that simply relates the lifetime to the imaginary part of the self energy $\Sigma''$.
\begin{equation}
\hbar/\tau [eV]= slope(\hbar\omega) [eV.\text{\normalfont\AA}] * \delta\nu~[1/\text{\normalfont\AA}]=Z * \Sigma''(\omega,T)
\label{Eq2}
\end{equation}
As we discussed in section III, the experimental dispersions are usually well described by renormalized LDA calculation, so that we can apply this procedure. It is straightforward when the dispersion is linear in the considered energy window (the two quantities are simply proportional), but would work as well for a parabolic \cite{KordyukCuprates} or arbitrary dispersion. In Fig. \ref{Largeur_MDC}, we plot the MDC linewidths as a function of $\hbar\omega$ and the slopes used for normalizing them.  The larger deviation from linearity is found for the $d_{xz}$/$d_{yz}$ electron band, which affects significantly the $\omega$ dependence of 1/$\tau$. However, we obtain a flat $\omega$ dependence for 1/$\tau$ (Fig. 1b) that echoes its flat temperature dependence (Fig. 1c) and gives confidence in this procedure. In LiFeAs, this normalization yields similar dependences for the hole and electron lifetimes on $d_{xz}$/$d_{yz}$, despite their very different dispersion, which again sounds correct. For $d_{xy}$, the $\omega$ dependences on the two pockets are indeed similar, except for an offset that we attribute to a better quality of the sample in which holes were measured.  

\vspace{3mm}
To go one step further, we fit the data in Fig. 1 to the $\omega^2$ or T$^2$ law expected for a Fermi liquid (Eq. 1), wherever it is possible. For $d_{xy}$, such fits describe well the data below $\sim$ 40~meV. This defines a coherent energy scale $E_{coh}$ below which the Fermi liquid is obeyed, as it is generally expected \cite{GeorgesRMP96}. We obtain $\gamma$=10 for LiFeAs and 16 for Co8. The difference in $\gamma$ is partly due to the more abrupt deviation of the $\omega^2$ behavior in Co8 and may not reflect essential physics (the broadening between $E_F$ and 80meV is for example nearly identical in the two compounds). The broadening of the $d_{xy}$ band with increasing temperature in Co8 can be described by the same exponent $\gamma$=16. For $d_{xz}$/$d_{yz}$, there is no clear FL regime, the data are dominated by a constant term. We could either fit a much smaller $\gamma$ (such as 2.5 in LiFeAs below 50~meV) or define a much smaller $E_{coh}$ ($\sim$10~meV in Co8). 

Recently, the lifetimes have been studied by optical spectroscopy both in BaFe$_2$As$_2$ doped with 10\% Co \cite{TytarenkoCondMat15} and LiFeAs \cite{DaiPRX15}. In both cases, the extracted lifetimes were found to follow the Fermi liquid behavior below 150K (i.e. $\pi$k$_B$T=40~meV) or $\hbar\omega\sim$50~meV. This $E_{coh}$ is in excellent agreement with that deduced from ARPES on $d_{xy}$. For 10\% Co, a value $\gamma$=4 was obtained from both energy and temperature dependence (a specific factor p should be added to Eq. 1 for the temperature dependence for optics \cite{StrickerPRL14,TytarenkoCondMat15}, which was estimated to be 1.5). For LiFeAs, $\gamma\sim7$ would be obtained from temperature dependence, if we assume the same p value. These values are smaller than those found here for $d_{xy}$, but would be compatible if averaged with the nearly flat dependence of $d_{xz}$/$d_{yz}$. This reinforces the idea that ARPES measures true QP lifetimes, but emphasizes that it is essential to consider the contribution of the two orbitals to interpret the results. 

For a simple metal, the renormalized bandwidth ZW would set a low energy scale. One expects $E_{coh}\ll$ ZW and $\gamma\sim$1/ZW \cite{GeorgesRMP96}. In iron pnictides, the situation is of course more complicated because of the interaction between the different bands, especially the coherence depends on the Hund\rq{}s couplings J between them \cite{WernerNatPhys12,deMediciPRL11}. Nevertheless, it seems important to keep in mind that ZW can be very small in iron pnictides, and, in fact, significantly smaller for $d_{xz}$/$d_{yz}$ than $d_{xy}$. Table I recalls that the bottom of the electron bands is located around 50~meV for $d_{xz}$/$d_{yz}$ and 150~meV for $d_{xy}$. The top of the hole bands are even closer from $E_F$, and closer for $d_{xz}$/$d_{yz}$ than $d_{xy}$. We are then in a situation where ZW is smaller for  $d_{xz}$/$d_{yz}$, despite the fact that Z can be larger for these orbitals. One could wonder whether it is possible at all to establish a Fermi liquid regime in such conditions. Indeed, the broad linewidth of $d_{xz}$/$d_{yz}$ could be interpreted as a totally incoherent behavior. Strictly speaking, QP are not defined if $\omega$<1/$\tau$, which is always the case here for $d_{xz}$/$d_{yz}$ (of course, this criterion is difficult to apply strictly for ARPES, due to the unknown impurity contribution). A very small $E_{coh}\sim$10~meV on  $d_{xz}$/$d_{yz}$ would explain the lack of  $\omega$ dependence, and the large $\gamma$$\sim$1/ZW, associated to it, could yield a high saturated value, larger than that of $d_{xy}$ near $E_F$. This explanation has the advantage to explain directly the similar behavior of Co8 and LiFeAs, where the relative ZW of $d_{xz}$/$d_{yz}$ and $d_{xy}$ are almost identical. However, it remains to be understood why the bands could be decoupled, so that the individual bandwidth becomes the relevant parameter.

\section{Conclusion}
We have demonstrated that ARPES is an efficient tool to resolve the properties of $d_{xy}$ and $d_{xz}$/$d_{yz}$ in iron pnictides. Especially, the two orbitals can be clearly separated on the electron pockets, which was rarely used, but allows accurate measurements and easy comparison between systems. Using both hole and electron pockets, we could confirm the increase of renormalization for $d_{xy}$ in LiFeAs compared to $d_{xz}$/$d_{yz}$. We further unveil an unexpected difference of behavior between the $d_{xy}$ and $d_{xz}$/$d_{yz}$ lifetimes, already present in Co8, where the renormalization are similar on the two orbitals. The $d_{xy}$ lifetime can be described by a Fermi liquid behavior below $E_{coh}\sim$40~meV with $\gamma\sim$10, while that of $d_{xz}$/$d_{yz}$ is nearly flat as a function of temperature or binding energy, implying (i) a smaller $\gamma$ or (ii) a smaller $E_{coh}$. 

(i) A smaller $\gamma$ could underline smaller correlations in $d_{xz}$/$d_{yz}$ compared to $d_{xy}$. In this case, based on our renormalization values, one would expect an evolution from nearly identical linewidth behaviors in Co8 to different behaviors in LiFeAs. As it is not the case, this explanation is not straightforward. Moreover, if $\gamma$ was smaller for $d_{xz}$/$d_{yz}$, there should be a much stronger impurity contribution to explain its larger linewidth at $E_F$. The possibility of orbital-dependent scattering to impurities is certainly an important point to consider when interpreting such data \cite{WadatiPRL10,Herbig15}. However, in Co-doped BaFe$_2$As$_2$, we observe similar broadenings of $d_{xz}$/$d_{yz}$ and $d_{xy}$ as a function of Co content, which rather suggests similar couplings. In fact, it is the $d_{xy}$ orbital that appears more sensitive to impurities, through temperature cycles in our measurements or Co doping in LiFeAs in ref \cite{YeFengPRX14}. Again, this does not support this scenario unambiguously. 

(ii) A smaller $E_{coh}$ on $d_{xz}$/$d_{yz}$ would explain at the same time the larger value at $E_F$ and the absence of well defined $\omega$ dependence.  We propose that the reduced effective bandwidth for $d_{xz}$/$d_{yz}$ ($\sim$50~meV) compared to $d_{xy}$ ($\sim$150~meV) could play a key role in defining $E_{coh}$. This would directly explain the similar behavior in Co8 and LiFeAs. We hope our data will stimulate further studies of how the Fermi liquid regime can be established in a multiband systems with small and very different effective bandwidths. 

Qualitatively, averaging the $d_{xy}$ and $d_{xz}$/$d_{yz}$ behavior revealed here reproduces quite well observations from optics or transport. However, our results suggest that it is the orbital content, rather than the hole or electron character, that is the main factor of differentiation. 

\vspace{3mm}

We thank A. van Roekeghem, S. Biermann and F. Rullier-Albenque for many useful discussions. This work benefitted from financial support from the ANR ‘‘Pnictides’’.

\bibliography{bib_test}

\end{document}